\documentclass[superscriptaddress,reprint,amssymb,amsmath,aps,pra]{revtex4-1}
\usepackage{graphics}
\usepackage{bbold}
\usepackage{graphicx}
\usepackage{bm}
\usepackage{bbm}
\usepackage{amsmath,amsthm}
\usepackage{lipsum}
\usepackage{amssymb}
\usepackage{color}

\begin{document}
	
	\title{Conditional states and entropy in qudit-qubit systems}
	
	\author{M. Bilkis}
	\affiliation{Instituto de F\'{\i}sica de La Plata, CONICET and Departamento de F\'{\i}sica,  
		Universidad Nacional de La Plata, C.C.\ 67, La Plata 1900, Argentina}
	\affiliation{F\'{\i}sica Te\`{o}rica: Informaci\'{o} i Fen\`{o}mens Qu\`{a}ntics, Departament de F\'{\i}sica, Universitat Aut\`{o}noma de Barcelona, 08193 Bellatera, Spain}
	\author{N. Canosa}
	\affiliation{Instituto de F\'{\i}sica de La Plata, CONICET and Departamento de F\'{\i}sica,  
		Universidad Nacional de La Plata, C.C.\ 67, La Plata 1900, Argentina}
	\author{R. Rossignoli}
	\affiliation{Instituto de F\'{\i}sica de La Plata, CONICET and Departamento de F\'{\i}sica,  
		Universidad Nacional de La Plata, C.C.\ 67, La Plata 1900, Argentina}
	\affiliation{Comisi\'on de Investigaciones Cient\'{\i}ficas de la Provincia de Buenos Aires (CIC), 
		La Plata 1900, Argentina}
	\author{N. Gigena}
	\affiliation{Instituto de F\'{\i}sica de La Plata, CONICET and Departamento de F\'{\i}sica,  
		Universidad Nacional de La Plata, C.C.\ 67, La Plata 1900, Argentina}
	
	\begin{abstract}
		We examine, in correlated mixed states of qudit-qubit systems, the set of all conditional qubit states 
		that can be reached after local measurements at the qudit based on rank-$1$  projectors.  While for a similar measurement at the qubit, 
		the conditional post-measurement qudit states lie on the surface of an ellipsoid, for a measurement at the qudit we show 
		that the set of post-measurement qubit states can form more complex solid regions. 
		In particular, we show the emergence, for some classes of mixed states, 
		of sets  which are the convex hull of solid ellipsoids and which may lead to cone-like and triangle-like shapes 
		in limit cases.  We also analyze the associated measurement dependent conditional entropy, providing a full analytic 
		determination of its minimum and of the minimizing local 
		measurement at the qudit for the previous states. Separable rank-$2$ mixtures are also discussed. 
	\end{abstract}
	\maketitle
	
	\section{Introduction}
	
	The study of quantum correlations and non-classical properties in composite quantum systems is of great current interest, 
	having deep implications in the field of quantum information  \cite{NC.00,Modi.12,ABC.16,FAC.10,dC.18}. A closely related 
	non-trivial problem is that of the determination of the set of post-measurement conditional states of one component after a remote local measurement on the other constituents. 
	In this context,  the concept of quantum steering ellipsoid   \cite{ve.02,Shi.11,Shi3.12,SS.13,Jev.14,Jev2.14,M.14,GR.14,Mc.17}, 
	also known as correlation ellipsoid \cite{SS.13,GR.14},  which denotes the set of all Bloch vectors to which one party could  
	collapse if the remote party were able to perform all possible measurements on  its side,  has provided a useful  geometric picture 
	in two-qubit  \cite{Shi.11,Shi3.12, SS.13,M.14,Jev.14,Jev2.14,GR.14,Mc.17,Hu.15} and also in  multi-qubit \cite{M.14,Cheng.16, exp.18} systems.  
	Recently, the experimental validation of the quantum steering ellipsoid for different two-qubit  states was reported \cite{exp.18}.
	
	This geometric approach has been important for understanding the measurement dependent conditional entropy and its minimizing measurement 
	\cite{Shi.11,Shi3.12,SS.13,GR.14,GRb.14}. This entropy measures the average conditional uncertainty in the post-measurement state of  
	the unmeasured constituent and its minimum plays a key role in the definition of the Quantum Discord \cite{OZ.01,HV.01,Modi.12,ABC.16,Lec.17,Bera.18}. 
	It is  also directly related with the entanglement of formation with a purifying third system \cite{KW.02}. The 
	steering ellipsoid has also provided necessary and sufficient conditions for the presence of  entanglement 
	in two-qubit  systems \cite{Jev.14,Jev2.14,M.14}, as well as strong  monogamy relations determined by  its volume in multi-qubit states \cite{M.14,Cheng.16,exp.18}. 
	The set of post-measurement reduced states in composite systems plays also a central role in the problem of quantum steering \cite{Sch1,WJD.07,SNC.14,K.15,Ga.15,CS.17,Bru.18}. 
	
	Most studies, however, have been concerned with measurements on a qubit component, where the set of all possible measurements 
	can be easily parameterized. In this work  we will consider instead measurements on a general qudit with dimension $d\geq 3$, 
	and analyze, for  mixed states of correlated qudit-qubit systems, the  set of the ensuing conditional states of  the unmeasured qubit 	after such measurements. 
	We first recall that while in bipartite pure states the conditional state of $B$ after a local measurement at the other system $A$ 
	(based on rank-$1$ local projectors)  is a pure state (which can be {\it any} pure state if the original global state is entangled 
	and the reduced state $\rho_B$ has full rank), 	in the case of mixed states the conditional state of the unmeasured system will be 
	in general mixed and lie within a certain subset of the full accessible space, 
	which is essentially determined 
	by the so-called correlation tensor of the global system
	\cite{ve.02,Jev.14,SS.13,GR.14,GRb.14} (see next section). 
	Such set may include from a pure state to the 
	maximally mixed state. And in the case of a qudit-qubit system, if such measurement is performed on the qubit, the set of conditional  
	states of the unmeasured qudit  forms  the surface of a three dimensional ellipsoid \cite{GR.14, GRb.14}. 
	
	Here we will show, however, that  for  measurements on the qudit, the set of conditional qubit states can form more complex 
	geometries, such as the convex hull of distinct solid ellipsoids and also  cone-like and  triangle-like shapes in limit cases, 
	providing analytic expressions.  We will also analyze the associated measurement dependent conditional entropy, providing general analytic 
	 results for its minimizing local measurement  at the qudit for certain classes of states, valid for general entropic forms,  together with their geometric picture. 
	
	We point out that  qudit-qubit systems admit several different physical realizations. In particular, it suffices to consider 
	the polarization degrees of freedom of a single photon as the qubit, while the qudit may correspond to its path degrees of freedom.  
	Both can be entangled through the use of beam displacers (as in \cite{exp.18})  or spatial light modulators (SLM)  
	(see for instance \cite{BL.14,KW.17,LR.19}).  Correlated qudit-qudit states can also be realized with two SPDC 
	(spontaneous parametric down conversion) photons using both polarization and the  transverse spatial correlations  
	 \cite{N.05,S.05,Pan.12}. And for qudits encoded in slit states generated through a SLM,  general measurements on the qudit can be realized, 
	 for instance, with the techniques described in   \cite{Sa.11}. Present results are then relevant for  determining the set of conditional 
	 polarization states that can be reached by measurements at the  spatial qudits  when the whole state is mixed.   
	Of course,  realizations of correlated qudit-qubit states through spin chains and arrays  are also feasible  
	(see for example \cite{Cho.08, senko.15, spin.18}).    
	
	The formalism is discussed in section II, where we  derive  analytic results for the  set of  conditional  qubit states after a 
	local measurement at the qudit for some classes of correlated qudit-qubit states. 
	In section III we examine the associated measurement determined conditional entropy, providing analytic results for its minimizing 
	measurement in the previous states.  We also include  results for general rank-$2$  separable states, with an application to mixtures 
	of aligned two-spin states of arbitrary spin. Conclusions are provided in section  IV.
	
	\section{Conditional states in bipartite systems}\label{II}
	\subsection{Formalism}
	\label{II.a}
	We first consider a general  bipartite system $A+B$, 
	with subsystem dimensions $d_A$ and $d_B$. We will use  orthogonal local operator bases formed by the identity 
	$\mathbbm{1}_S$ plus $d_S^2-1$ hermitian traceless operators $\bm{\sigma}_{S}$ satisfying 
	\begin{equation}\text{Tr} \, 
	[\sigma_{S\mu} \sigma_{S\nu}] = d_S \delta _{\mu \nu}\,,\;\;\;S=A,B\,.\label{ort}\end{equation} 
	A  general mixed state $\rho_{AB}$ can then be written as  
	\cite{F.83} 
	\begin{equation} \label{eq:fanobloch}
	\rho_{AB} = \rho_A\otimes\rho_B+ \frac{1}{ d_A d_B}\sum_{\mu,\nu} C_{\mu \nu}  \sigma_{A\mu} \otimes \sigma_{B\nu},
	\end{equation}
	where $\rho_{A(B)} = {\rm Tr}_{B(A)}\rho_{AB}$ are the reduced states 
	\begin{equation} 
	\rho_{S} =\frac{1}{d_S} (\mathbbm{1}_S+\bm{r}_S \cdot \bm{\sigma}_{S})\,, \label{rhos}
	\end{equation}
	with  
	$\bm{r}_{S}=\langle \bm{\sigma}_{S}\rangle= {\rm Tr}_S\,[\rho_{S}\,\bm{\sigma}_{S}]$, while  
	\begin{equation}
	C_{\mu\nu} =\langle \sigma_{A\mu}\otimes\sigma_{B\nu}\rangle-\langle \sigma_{A\mu}\rangle\langle\sigma_{B\nu}\rangle \label{cten}
	\end{equation}
	are the elements of the {\it correlation tensor} \cite{GR.14}.  
	
	We now assume that a local 
	measurement based on rank-$1$  local projectors is performed on  side $A$. 
	It is worth mentioning that 
	this type of POVM is sufficient to minimize 
	the measurement dependent conditional entropy \cite{Modi.12,GRb.14}. Moreover,  they allow for analytical expressions in the case of simple entropic forms \cite{GRb.14}  (see sec.\ \ref{III.C} for further details). The projectors can be expressed as 
	\begin{equation}
	\Pi^{A}_{\bm k} = |\bm{k}_A\rangle\langle \bm{k}_A|=\frac{1}{d_A}(\mathbbm{1}_A + \bm{k} \cdot \bm{\sigma}_A), \end{equation}
	with $\bm{k}=\langle\bm{k}_A|\bm{\sigma}^A|\bm{k}_A\rangle$  a  vector satisfying $|\bm{k}|^2=d_A-1$. 
	
	The conditional  post-measurement  state of $B$ is 
	\begin{eqnarray}
	\rho_{B/\bm{k}}&=&\,p_{\bm{k}}^{-1}{\rm Tr}_A[\rho_{AB}\, \Pi^A_{\bm{k}}\otimes \mathbbm{1}_B]=
	p_{\bm{k}}^{-1}  \langle\bm{k}_A|\rho_{AB}|\bm{k}_A\rangle\nonumber\\&=&
	\frac{1}{d_B}[\mathbbm{1}+\bm{r}_{B/\bm{k}}
	\cdot\bm{\sigma}_B], \label{cqu}
	\end{eqnarray}
	where $p_{\bm{k}}={\rm Tr}_A[\rho_{A}\Pi^A_{\bm{k}}]=\frac{1}{d_A}(1+\bm{r}_A\cdot\bm{k})$ is the probability of measuring state $|\bm{k}_A\rangle$ and 
	\begin{equation} 
	\bm{r}_{B/\bm{k}} ={\rm Tr}_B[\rho_{B/\bm{k}} \bm{\sigma}_B]= \bm{r}_B + \frac{C^T\bm{k}}{1 + \bm{r}_A\cdot \bm{k}}\,, \label{rac}
	\end{equation}
	is the conditional average of $\bm{\sigma}_B$ after  result $\bm{k}$ at $A$.  
	
	The complete local measurement will be defined by a set of $m$  operators $M_j^A=\sqrt{r_j}\,\Pi_{\bm{k}_j}^A$, $r_j>0$, 
	satisfying $\sum_j (M_j^A)^\dagger M_j^A=\sum_j r_j\Pi_{\bm{k}_j}^A=\mathbbm{1}_A$, i.e.,  
	\begin{equation}
	\sum_j r_j =d_A \;,\;\;\;\sum_j r_j \bm{k}_j=\bm{0}\,.\label{condg}
	\end{equation} 
	The probability of result $j$ is  $p_j=r_j p_{\bm{k}_j}$,   with (\ref{condg}) ensuring $\sum_j p_j=1$ and   
	$\sum_j p_j \bm{r}_{B/\bm{k}_j}=\bm{r}_B$, i.e.\, $\sum_j p_j \rho_{B/\bm{k}_j}=\rho_B$, 
	preventing faster than light signalling from $A$ to $B$. 
	Standard von Neumann measurements correspond to $m=d_A$, $r_j=1$ $\forall$ $j$ and $\Pi_{\bm{k}_j}^A$
	orthogonal projectors. 
	
	If $\rho_{AB}$ has  local support on a certain subspace ${\cal S}_A$ of $A$ of dimension $d'_A<d_A$, 
	we can always write
	\begin{equation}|\bm{k}_A\rangle=\sqrt{q}|\bm{k}^{\parallel}_A\rangle+\sqrt{1-q}|\bm{k}^\perp_A\rangle\,,\label{kq}\end{equation} 
	where $|\bm{k}^{\parallel}_A\rangle\in{\cal S}_A$,  $q\in[0,1]$ and $|\bm{k}^\perp_A\rangle$ is orthogonal to ${\cal S}_A$,  
	such that $\langle \bm{k}_A|\rho_{AB}|\bm{k}_A\rangle=q\langle \bm{k}^{\parallel}_A|\rho_{AB}|\bm{k}^{\parallel}_A\rangle$. 
	Hence, for conditional states the effects of a complete local measurement based on operators $M_j^A=\sqrt{r_j}\,\Pi^A_{\bm{k}_j}$ 
	are the same as those of a measurement in the subspace ${\cal S}_A$ based on  operators $M'_j=\sqrt{q_j r_j}\,\Pi^A_{\bm{k}_j^{\parallel}}$, 
	satisfying  $\sum_j {M'_j}^\dagger M'_j=\mathbbm{1}_{d'_A}$. 
	
	In what follows we will consider a {\it qudit-qubit} system ($d_B=2$). We will analyze the whole set of post-measurement conditional qubit  
	states $\rho_{B/\bm{k}}$ of $B$,  characterized by the now Bloch vectors (\ref{rac}) ($\bm{\sigma}_B$ are now the standard Pauli operators), 
	that can be reached for any possible rank-$1$  projector $\Pi^A_{\bm{k}}$.  We first recall that for similar measurements at the qubit, 
	 the set of all conditional postmeasurement vectors $\bm{r}_{A/\bm{k}}$ of qudit $A$ forms the surface of a three-dimensional 
	ellipsoid (for a rank-$3$ correlation tensor)  \cite{ve.02,GR.14}, whose semiaxes are determined by the correlation tensor $C$ and the vector $\bm{r}_B$.
	In contrast, for a measurement on the qudit $A$, we will  show that the set of all post-measurement qubit Bloch vectors $\bm{r}_{B/\bm{k}}$ 
	will be in general a region with finite volume, which may have shapes more general than a single ellipsoid. 
	
	\subsection{Mixture of a pure state with the maximally mixed state} 
	\label{II.b}
	As a first example, we consider the qudit-qubit state 
	\begin{equation}
	\rho_{AB}=p|\Psi\rangle\langle\Psi|+(1-p)\frac{\mathbbm{1}_{AB}}{2d_A}\,,
	\label{rh1}
	\end{equation}
	where $|\Psi\rangle\equiv|\Psi_{AB}\rangle$ stands for a general pure state and $p\in[0,1]$.  Positivity of $\rho_{AB}$ is, 
	nonetheless, ensured for $-\frac{1}{2d_A-1}\leq p\leq 1$, with negative  values of $p$ representing depletion of state $|\Psi\rangle$ from the maximally mixed state.  
	
	By means of the Schmidt decomposition of $|\Psi\rangle$, we may always choose orthogonal local states $|0_{A(B)}\rangle$ and $|1_{A(B)}\rangle$ 
	such that $|\Psi\rangle $ can be written as 
	\begin{equation}|\Psi\rangle=\cos\tfrac{\beta}{2}|00\rangle+\sin\tfrac{\beta}{2}|11\rangle\,,
	\label{psi}
	\end{equation}
	where  $|ij\rangle=|i_A\rangle\otimes|j_B\rangle$ and  
	the angle $\beta\in[0,\pi/2]$ is determined by its concurrence \cite{WW.97} ($|\Psi\rangle$ can be seen as an effective two-qubit state)  $C_{AB}=\sin\beta$.  
	The mixed state (\ref{rh1}) has then non-positive partial transpose \cite{Pe.96, HHH.96} (i.e., positive negativity \cite{VW.02}) 
	for $|p|\sin\beta>(1-p)/d_A$, i.e.\  $p>(1+d_A\sin\beta)^{-1}>0$.  
	
\begin{figure}
		\begin{center}
			\hspace*{-.5cm}\includegraphics[width=0.5\textwidth]{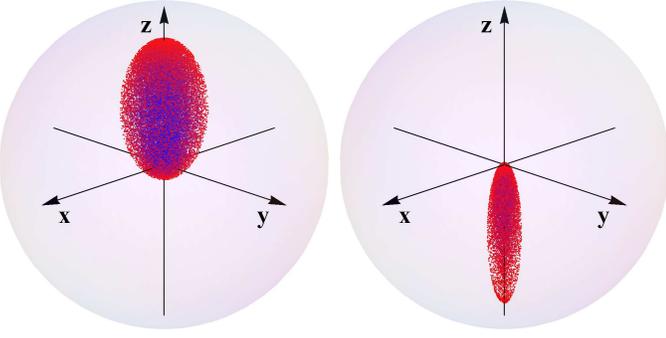}
		\end{center}
		\caption{Left: The set of conditional post-measurement states of  qubit B, represented by the Bloch vectors  (\ref{pol})  within the qubit Bloch sphere, after random measurements on the qudit A based on rank-$1$ projectors, for the qudit-qubit mixed state (\ref{rh1}) with $p=0.5$, $\beta=\pi/10$ and $d_A=4$ 
			(for which $\rho_{AB}$ is entangled). The set forms a filled ellipsoid if $p<1$,  with the origin at one of its foci 
			and eccentricity determined  by $p$ and the entanglement of $|\Psi\rangle$   Eq.\ (\ref{e}), becoming a sphere 
			when maximally entangled. The color indicates the value of $q$ in (\ref{pw})--(\ref{pol})  
		    (red (blue) for high (low)  $q$), which determines the weight of the pure term 
		    in the post-measurement qubit state (\ref{rbk1a}). 
			Right: An ellipsoid with opposite orientation is  obtained for feasible ($p\geq-\frac{1}{2d_A-1}$) 
			negative values of $p$ in (\ref{rh1}), here  shown  for $p=-0.14$.} \label{f1}
	\end{figure}

	We now show that for $p<1$ and $d_A\geq 3$,  the set of all possible conditional qubit states $\rho_{B/\bm{k}}$ after a  
	projective measurement at the qudit with result $\bm{k}$,   will be 
	{\it a filled ellipsoid symmetric around the local $z$ axis, with the origin as one of its foci and the major semiaxis 
		along $z$} (Fig.\ \ref{f1} left). The local $z$ axes are, of course, those defined by the Schmidt states, 
	such that $\sigma_{Sz}=|0_S\rangle\langle 0_S|-|1_S\rangle\langle 1_S|$. 
	
	{\it Proof:} We first note that if the state $|\bm{k}_A\rangle$ of  $\Pi^A_{\bm{k}}$ is restricted to the subspace spanned by the states $\{|0_A\rangle,|1_A\rangle\}$, 
	the situation is similar to that of a two-qubit system and the surface of an ellipsoid will be obtained \cite{GR.14}. 
	But if $|\bm{k}_A\rangle$ has just a component within this subspace, 
	$\bm{r}_{B/\bm{k}}$ will lie within the previous ellipsoid due to the smaller overlap with the state $|\Psi\rangle$,  
	filling the whole ellipsoid as this component diminishes. 
	We now prove this result explicitly, providing the ellipsoid parameters. 
	Discarding a global phase, 
	a general pure qudit state $|\bm{k}_A\rangle$  can be here written as (Eq.\ (\ref{kq})) 
	\begin{eqnarray}
	|\bm{k}_A\rangle&=&\sqrt{q}|\bm{k}_A^{\parallel}\rangle+\sqrt{1-q}\,|\bm{k}_A^\perp\rangle,
	\;\nonumber\\
	|\bm{k}_A^{\parallel}\rangle&=&\cos\tfrac{\alpha}{2} |0_A\rangle+
	\sin\tfrac{\alpha}{2}e^{-i\phi}|1_A\rangle\,,
	\label{pw}\end{eqnarray}
	with $q\in[0,1]$, $\alpha\in[0,\pi]$, $\phi\in[0,2\pi)$ and $|\bm{k}_A^{\perp}\rangle$ orthogonal to  
	$|0_A\rangle$ and $|1_A\rangle$. The normalized  conditional postmeasurement 
	qubit state (\ref{cqu}) becomes  
	\begin{eqnarray}\rho_{B/\bm{k}}&=&
	p_{\bm k}^{-1}\big[pq\langle \bm{k}_A^{\parallel}|\Psi\rangle\langle\Psi|\bm{k}_A^{\parallel}\rangle+\frac{1-p}{2d_A}\mathbb{1}_B\big] \,,\label{rbk1a}
	\end{eqnarray}  
	with $\langle\bm{k}_A^{\parallel}|\Psi\rangle=
	\cos\tfrac{\alpha}{2}
	\cos\tfrac{\beta}{2}|0_B\rangle+e^{i\phi}\sin\tfrac{\alpha}{2}
	\sin\tfrac{\beta}{2}|1_B\rangle$ and 
	$p_{\bm{k}}=\tfrac{1}{2}pq(1+
	\cos\beta\cos\alpha)
	+\frac{1-p}{d_A}$. 
	
	The ensuing Bloch vector 
	$\bm{r}_{B/\bm{k}}={\rm Tr}\,[\rho_{B/\bm{k}}\,\bm{\sigma}]$ is   
	\begin{eqnarray}
	\bm{r}_{B/\bm{k}}&=&\frac{pq}{2p_{\bm{k}}}
	(\sin\alpha\sin\beta\cos\phi,
	\sin\alpha\sin\beta\sin\phi,\cos\alpha+\cos\beta)
	\nonumber\\
	&=&r_{B/\bm{k}}(\sin\theta\cos\phi,\sin\theta\sin\phi,\cos\theta)\,,\label{pol}
	\end{eqnarray}
	where (\ref{pol}) is its {\it polar representation}, with 
	$\cos\theta=\frac{\cos\alpha+\cos\beta}{1+\cos\alpha\cos\beta}$,  and  $r_{B/\bm{k}}$  
	given by 
	\begin{equation}
	r_{B/\bm{k}}=\frac{a(1-e^2)}{1-e\cos\theta}
	\label{elips},
	\end{equation}
	with  $\theta\in[0,\pi]$ if $\alpha\in[0,\pi]$ (and $\beta\neq 0$). Here  
	\begin{eqnarray}
	e&=&\frac{(1 - p)\cos\beta}{1 - p + \frac{1}{2}p q d_A\sin^2\beta}\,,\label{e}\\
	a&=&\frac{pq(\frac{1-p}{d_A}+\frac{1}{2}pq\sin^2\beta)}{2\Delta}\,, \label{a}
	\end{eqnarray}
	with  $\Delta=(\frac{1-p}{d_A}+pq\sin^2\frac{\beta}{2})(\frac{1-p}{d_A}+pq\cos^2\frac{\beta}{2})$.  
	
	Thus, at fixed $\beta>0$ and $q>0$, Eq.\ (\ref{elips}) represents an ellipsoid symmetric around the $z$ axis 
	with {\it eccentricity $|e|$ and major semiaxis of length $|a|$ along $z$}, with the origin at its focus. 
	All ellipsoids are enclosed within that for $q=1$, for which $|a|$ is maximum.  Thus, for $d_A\geq 3$, 
	variation of $q$ in the interval $[0,1]$ leads to a {\it filled ellipsoid}, as shown in Fig.\ \ref{f1}. 
	In the qubit case $d_A=2$, $q=1$ and just its surface remains. \qed
	
	In cartesian coordinates, the ellipsoid equation reads   \begin{equation} \frac{x^2+y^2}{b^2}+\frac{(z-z_c)^2}{a^2}=1,
	\end{equation}
	where  $b=a\sqrt{1-e^2}=pq\sin\beta/(2\sqrt{\Delta})$ is the minor semiaxis length, 
	and $z_c=a e=pq\frac{1-p}{2d_A\Delta}\cos\beta$ the  $z$-coordinate of its center. Let us now verify some limit cases. 
	For $\beta=\pi/2$, $|\Psi\rangle$ is maximally entangled and $e=0$: The filled ellipsoid  becomes a filled sphere  
	of radius $a=b=[1+\frac{2(1-p)}{pd_A}]^{-1}$, 
	centered at the origin ($z_c=0$). 
	On the other hand, for  $\beta=0$,  $|\Psi\rangle$ becomes a product state and $e=1$, implying that the ellipsoid reduces 
	to a {\it segment}  along the $z$ axis ($b=0$, $z_c=a$), starting at the origin and ending at $2a=[1+\frac{1-p}{pqd_A}]^{-1}$. 
	
	Finally, in the pure state limit $p=1$, the set of post-measurement states becomes  the whole Bloch sphere {\it surface} (it is always pure) 
	for {\it any} entangled $|\Psi\rangle$, as $e\rightarrow 0$ and $a\rightarrow 1$ for any $\beta\in(0,\pi)$, 
	while for a separable $|\Psi\rangle$ ($\beta=0$) it obviously reduces to the point $(1,0,0)=\bm{r}_B$.   
	If the qudit dimension $d_A$ increases (at fixed $\beta$ and $p\in (0,1)$),  $a$ and hence the volume of the ellipsoid increases, 
	whereas the eccentricity 
	decreases. For $pqd_A\gg 1$, $a\approx 1-\frac{2(1-p)}{pqd_A\sin^2\beta}$, 
	$e\approx \frac{2(1-p)\cos\beta}{pqd_A\sin^2\beta}$. 
	
	We also note that Eqs.\ (\ref{elips})--(\ref{a}) remain  valid for {\it negative} values $\frac{-1}{2d_A-1}\leq p< 0$, 
	in which case $a$ and $z_c$ change sign.  Hence, an ellipsoid also follows from a state $\rho_{AB}$ maximally mixed in the 
	$2d_A-1$-dimensional subspace orthogonal to $|\Psi\rangle$ (Fig.\ \ref{f1} right).
	
	\subsection{Mixture of two pure states with the maximally mixed state\label{II.c}}
\begin{figure}
		\begin{center}
			\includegraphics[width=0.32\textwidth]{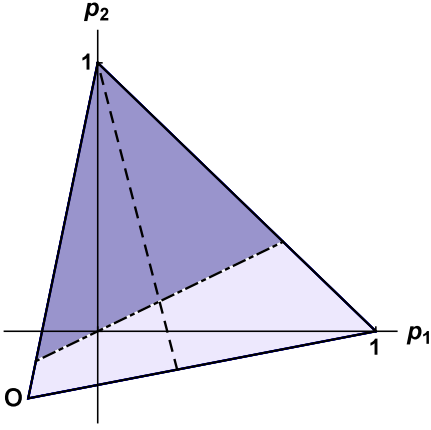}
		\end{center}
		\caption{The triangle in  $(p_1,p_2)$ space corresponding to a physical state in the qudit-qubit mixed state (\ref{rh2}).  
			{\bf O} indicates the  vertex at 
			$\frac{-1}{2 d_A - 2}(1,1)$.  In the  case  $\beta_1=\pi/2$, $\beta_2=0$,
			points $(p_1, p_2)$ on the right of the dashed  line correspond to an entangled state with non-zero negativity,  
			while those above (below) the dashed-dotted line, i.e.\ dark (light) colored  sectors, lead to a filled ``ice-cream'' (ellipsoid) shape of 
			the set of post-measurement qubit states (see Eq.\ (\ref{cv})).}
		\label{f2}
	\end{figure}

	For other states $\rho_{AB}$, the set of post-measurement qubit states may adopt more complex forms. Let us consider, for instance, the  state 
	\begin{equation}
	\rho_{AB}=p_1|\Psi_1\rangle\langle\Psi_1|+p_2|\Psi_2\rangle\langle\Psi_2| +p_0\frac{\mathbbm{1}_{AB}}{2d_A}\,,\label{rh2}
	\end{equation}
	where $p_0=1-p_1-p_2$. 
	Condition $\rho_{AB}\geq 0$ implies $p_1 +p_2\leq 1$,  $p_1-(2d_A-1)p_2\leq 1$ and $p_2-(2d_A-1)p_1\leq 1$,  
	which  delimit a triangle in the $(p_1,p_2) $ plane with vertices at $(p_1,p_2)=(1,0)$, $(0,1)$ and 
	$\frac{-1}{2d_A-2}(1,1)$ (Fig.\ \ref{f2}). 	We will focus on the case where 
	the states $|\Psi_1\rangle$ and $|\Psi_2\rangle$ are orthogonal and have orthogonal supports at the qudit side.
	
		\subsubsection{Two entangled states} 
 We first consider the case where these states are entangled, such that their Schmidt decompositions are 
	\begin{eqnarray}|\Psi_1\rangle&=&\cos\tfrac{\beta_1}{2}|00_1\rangle+\sin\tfrac{\beta_1}{2}|11_1\rangle\,,
	\label{psi1}\\
	|\Psi_2\rangle&=&\cos\tfrac{\beta_2}{2}|20_2\rangle+\sin{\tfrac{\beta_2}{2}}|31_2
	\rangle \,,\label{psi2}
	\end{eqnarray}
	where qudit states $|i_A\rangle$, $i=0,1,2,3$, are all orthogonal while qubit states $|j_{1_B}\rangle$ are not necessarily 
	orthogonal to $|j'_{2_B}\rangle$.  We are assuming here $d_A\geq 4$.  
	The partial transpose will have a negative eigenvalue associated with  state $|\Psi_i\rangle$ if 
	$p_i \sin\beta_i> p_0/d_A$, i.e.\, $p_1>\frac{1-p_2}{1+d_A\sin\beta_1}$ for $|\Psi_1\rangle$ and $p_2>\frac{1-p_1}{1+d_A\sin\beta_2}$ 
	for  $|\Psi_2\rangle$, having at most two negative eigenvalues. 
	
\begin{figure}
		\hspace*{-0.4cm}\includegraphics[width=0.48\textwidth]{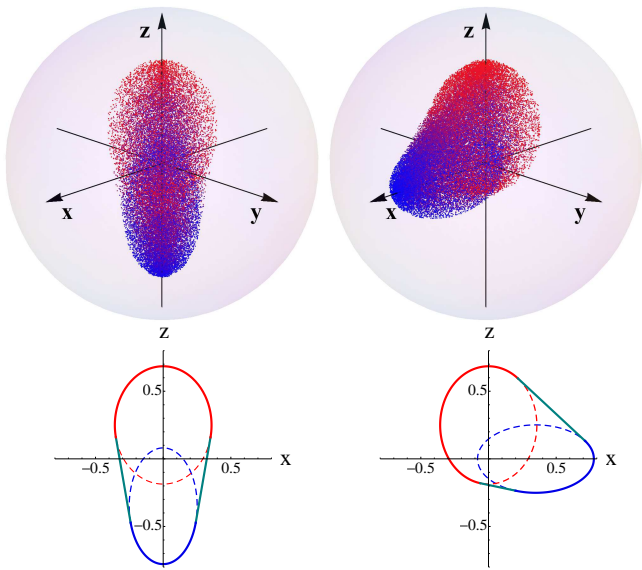}
		\caption{The set of post-measurement qubit states after random  measurements on the 
			qudit for the state (\ref{rh2}). It is the convex hull of the ellipsoids  (\ref{eli}) associated with each state $|\Psi_i\rangle$ 
			of the mixture, as indicated in the lower panels, which  depict the set boundaries in the $xz$ plane (solid lines), 
			together with the ellipsoids projection (solid and dashed lines). 
			The parameters are $p_1=0.2$, $\beta_1=\pi/5$, $p_2=0.3$, $\beta_2=\pi/10$ and $d_A=6$, for which $\rho_{AB}$ is entangled, 
			with the major semiaxes of the ellipsoids in opposite (left) and orthogonal (right) directions. The color indicates 
			the relative weight of each state $|\Psi_i\rangle$ in the final state (\ref{rg}) (the value of $q$ in Eq.\ (\ref{kag})), 
			with red (blue) denoting prominence of $|\Psi_1\rangle$ ($|\Psi_2\rangle$).}
		\label{f3}
	\end{figure}

	The set of reachable post-measurement qubit states for measurements based on rank-$1$  local projectors at the qudit, 
	will be essentially the convex hull of the ellipsoids associated with each state, as seen in Fig.\ \ref{f3}.  
	A general pure state $|\bm{k}_A\rangle$ of the qudit can now be written as 
	\begin{equation}|\bm{k}_A\rangle=
	\sqrt{q'}(\sqrt{q}|\bm{k}_{1A}\rangle+e^{-i\psi}\sqrt{1-q}|\bm{k}_{2A}\rangle)+\sqrt{1-q'}\,|\bm{k}_A^{\perp}\rangle\,,\label{kag}
	\end{equation}
	where $q,q'\in[0,1]$, $|\bm{k}_{A}^{\perp}\rangle$ is orthogonal to all  $|i_A\rangle$  with  $i\leq 3$ (if $d_A\leq 4$ we set $q'=1$), and 
	\begin{equation}
	\begin{array}{lcl}
	|\bm{k}_{1A}\rangle&=&\cos\tfrac{\alpha_1}{2}|0_A\rangle+e^{-i\phi_1}
	\sin\tfrac{\alpha_1}{2}|1_A\rangle\,,\\
	|\bm{k}_{2A}\rangle&=&
	\cos\tfrac{\alpha_2}{2}|2_A\rangle+e^{-i\phi_2}\sin\tfrac{\alpha_2}{2}|3_A\rangle\,.
	\end{array}
	\label{kia}
	\end{equation}
	
	The resulting conditional state of the qubit is  
	\begin{eqnarray}\rho_{B/\bm{k}}&=&p_{\bm{k}}^{-1}\big\{
	q'\big[qp_1\langle \bm{k}_{1A}|\Psi_1\rangle\langle\Psi_1|\bm{k}_{1A}\rangle\nonumber\\
	&&+(1-q)p_2\langle \bm{k}_{2A}|\Psi_2\rangle\langle\Psi_2|\bm{k}_{2A}\rangle\big]
	+\frac{p_0}{2d_A}\mathbbm{1}_B\big\}\label{rg}\;\;\;\\
	&=&\frac{q\, p_{\bm{k}_1}}{p_{\bm{k}}}\rho_{B/\bm{k}_1}+
	\frac{(1-q)\, p_{\bm{k}_2}}{p_{\bm{k}}}\rho_{B/\bm{k}_2}
	\,,\label{rhk12}
	\end{eqnarray}
	where $\langle \bm{k}_{iA}|\Psi_i\rangle=\cos\tfrac{\alpha_i}{2}
	\cos\tfrac{\beta_i}{2}|0_i\rangle+e^{i\phi_i}\sin\tfrac{\alpha_i}{2}
	\sin\tfrac{\beta_i}{2}|1_i\rangle$,  $p_{\bm{k}}=q\, p_{\bm{k}_1}+(1-q)\,p_{\bm{k}_2}$ and 
	\begin{eqnarray}\rho_{B/\bm{k}_i}&=&
	\frac{1}{p_{\bm{k}_i}}\big(q'p_i\langle\bm{k}_{iA}|\Psi_i\rangle\langle\Psi_i|\bm{k}_{iA}\rangle+\frac{p_0}{2d_A}\mathbbm{1}_B\big),\label{rhi}\\
	p_{\bm{k}_i}&=&q'p_i(1+\cos\beta_i\cos\alpha_i)/2+p_0/d_A\,,
	\end{eqnarray} 
	are the conditional post-measurement qubit states and probabilities obtained for 
	$q=1$ $(i=1)$ and $0$ ($i=2$). 
	
	Hence, the post-measurement state (\ref{rhk12}) is just the {\it convex combination} of the states (\ref{rhi}), 
	with $q p_{\bm{k}_1}/p_{\bm{k}}$ covering all values between $0$ 
	and $1$ as $q$ varies from $0$ to $1$ (even for negative values of $p_i$ provided $\rho_{AB}$ is positive and $q'p_1p_2\neq 0$). 
	The Bloch vector obtained from (\ref{rg}) can then be written as 
	\begin{equation}
	\bm{r}_{B/\bm{k}}=\frac{q\, p_{\bm{k}_1}}{p_{\bm{k}}}\bm{r}_{B/\bm{k}_1}+
	\frac{(1-q)\, p_{\bm{k}_2}}{p_{\bm{k}}}\bm{r}_{B/\bm{k}_2}\label{rt}
	\end{equation}
	where $\bm{r}_{B/\bm{k}_i}$ are the vectors determined by $\rho_{B/\bm{k}_i}$:  
	\begin{eqnarray}\bm{r}_{B/\bm{k}_i}&=&\frac{q'p_i}{2p_{\bm{k}_i}}
	R_i(\sin\alpha_i\sin\beta_i\cos\phi_i,
	\sin\alpha_i\sin\beta_i\sin\phi_i,\nonumber\\&&\cos\alpha_i+\cos\beta_i)\nonumber\\
	&=&r_{B/\bm{k}_i}R_i(\sin\theta_i\cos\phi_i,\sin\theta_i\sin\phi_i,\cos\theta_i)\,.\label{rki}
	\end{eqnarray}
	Here $R_i$ are the operators rotating the 
	original $z$ axis to the $z_i$ axis determined by the states $|0_{i}\rangle$, $|1_i\rangle$, 
	 while  $\cos\theta_i=\frac{\cos\alpha_i+\cos\beta_i}{1+\cos\alpha_i\cos\beta_i}$ and 
	\begin{eqnarray}
	r_{B/\bm{k}_i}&=&\frac{a_i(1-e_i^2)}{1-e_i\cos\theta_i}\,,\label{eli}\end{eqnarray}
	with 
	\begin{eqnarray}
	e_i&=&\frac{p_0\cos\beta_i}{p_0+\frac{1}{2}p_i q' d_A\sin^2\beta_i}\,,\\
	a_i&=&\frac{p_iq'(p_0/d_A+\frac{1}{2}p_i q'\sin^2\beta_i)}{2(p_0/d_A+p_iq'\cos^2\frac{\beta_i}{2})(p_0/d_A+p_i q'\sin^2\frac{\beta_i}{2})}\,.\;\;\;
	\end{eqnarray}
	Therefore, the ensuing set of post-measurement vectors obtained for all values of $\alpha_1,\phi_1,\alpha_2,\phi_2,q$ 
	and $q'$ in (\ref{kag})--(\ref{kia}) will be the {\it convex hull of the filled ellipsoids (\ref{rki}) determined by  $\bm{r}_{B/\bm{k}_i}$}. 
	All sets will be contained within that obtained for $q'=1$, entailing that the  set can be obtained by setting $q'=1$ 
	(and varying all other measurement parameters). The same set is  then also obtained for $d=4$ (where $q'=1$). Present results 
	can be straightforwardly extended to a mixture of several pure states $|\Psi_i\rangle$ with orthogonal supports at $A$. 
	
	As illustration, Fig.\ \ref{f4} depicts the resulting figure when both ellipsoids have colinear  (left) or orthogonal (right) major semiaxes, for $d_A=6$. 
	We note that the ensuing $\rho_{AB}$ is entangled, with two negative eigenvalues of the partial transpose. 
	
	The condition which ensures that the major semiaxes of the second ellipsoid will protrude above the first ellipsoid surface is just 
	\begin{equation}
	p_2\geq\frac{p_1\sin^2\beta_1}{(1+\cos\beta_2)(1-\cos\beta_1\cos\gamma)}
	\label{ptr1}\,,
	\end{equation}
	where $\gamma$ is the angle between both major semiaxes and we have assumed $p_i\geq 0$ (for negative values the inequality should be inverted). 
	
	\subsubsection{Limit cases \label{IID}}
	{\it ``Ice-cream''} shapes. 
	We now examine the case where one  of the states $|\Psi_i\rangle$ is separable. In this case we can consider $d_A\geq 3$. 
	When $\beta_2=0$,  $|\Psi_2\rangle$ in (\ref{rh2})  becomes {\it separable}: 
	\begin{equation}
	|\Psi_2\rangle=|2_A\rangle|0_{2}\rangle\,,\;\;
	|0_{2}\rangle=\cos\tfrac{\gamma}{2}|0_1\rangle+e^{i\eta}
	\sin\tfrac{\gamma}{2}|1_1\rangle\,,
	\label{cone}
	\end{equation}
	where $|0_{2}\rangle$ is  an arbitrary qubit state.  
	The second ellipsoid then reduces to a  {\it segment}: We obtain
	\begin{equation}
	\bm{r}_{B/\bm{k}_2}=\frac{q'p_2\cos^2\tfrac{\alpha_2}{2}}{q'p_2\cos^2\tfrac{\alpha_2}{2}+p_0/d_A}
	\langle 0_2|\bm{\sigma}|0_2\rangle
	\end{equation}
	with $\langle 0_2|\bm{\sigma}|0_2\rangle=
	(\sin\gamma\cos\eta,\sin\gamma\sin\eta,\cos\gamma)$.  Therefore, when  varying $q'$ and/or $\alpha_2$, 
	it leads to a segment linking the origin with the vector 
	\begin{equation} \bm{r}_v=\frac{p_2(\sin\gamma\cos\eta,\sin\gamma\sin\eta,\cos\gamma)}{p_2+p_0/d_A}\,.\label{rv}
	\end{equation}
	By conveniently choosing the $x$ axis we may obviously always set $\eta=0$.  In the qutrit case $d_A=3$, 
	$q'=1$ and $\alpha_2=0$, so that here  $\bm{r}_{B/\bm{k}_2}=\bm{r}_v$.  
	
	The final set obtained after covering all values of  $q\in[0,1]$ and  $\alpha_1,\phi_1$ will lead  to the {\it convex hull 
		of the first ellipsoid and the segment ending in $\bm{r}_v$} (or equivalently, the point $\bm{r}_v$). The parameter $q'$ will have no effect 
	in the full final set, since the variation of $q$ will already produce a filled volume, so that this result is also valid for $d_A=3$.  
	
\begin{figure}\hspace*{-1.5cm}\includegraphics[width=0.48\textwidth]{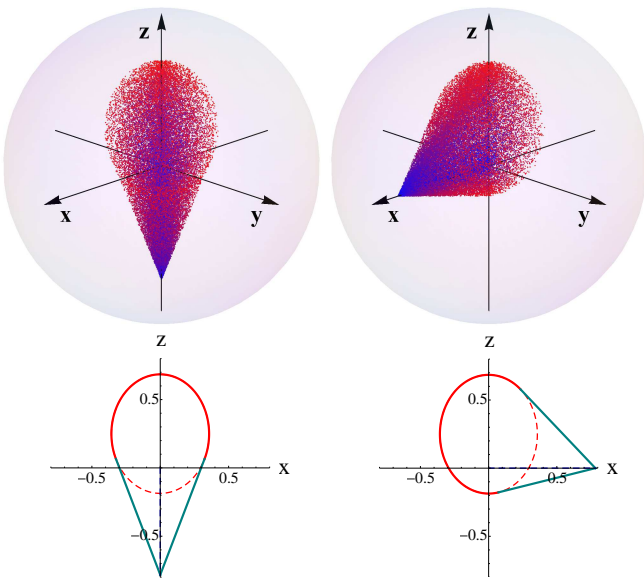}
		\caption{The set of post-measurement qubit states after random  measurements on the qudit, 
			for the mixture  (\ref{rh2}) with an entangled state $|\Psi_1\rangle$ ($\beta_1=\pi/5$) and  a {\it separable} state $|\Psi_2\rangle$ ($\beta_2=0$). 
			The resulting set is a cone topped with an ellipsoid (``ice-cream'' shape), with
			the cone vertex $r_v$ determined by the separable state $|\Psi_2\rangle$ (Eq.(\ref{rv})). The plots correspond to $r_v$ in the negative $z$ axis (left) 
			and positive $x$ axis (right), with $p_1=0.2$, $p_2=0.3$ and $d_A=6$. The lower panels depict the formation of the cone as a result of the convex hull of the ellipsoid and the point $\bm{r}_v$.}    
		\label{f4}
	\end{figure}

	If $\bm{r}_v$ lies within the ellipsoid, the final set will still be a filled ellipsoid. However, when it lies {\it outside},  
	the final figure will be a {\it cone with vertex at $\bm{r}_v$ topped with the ellipsoid}, 
	with the cone straight borders ending {\it tangent} to the ellipsoid surface. This leads to an ``{\it ice-cream}''-like shape, as seen in 
	Fig.\ \ref{f4} for the cases where the segment is collinear (left panel) or orthogonal (right panel) to the major semiaxis of the ellipsoid. 
	
	From Eq.\ (\ref{ptr1}) for $\beta_2=0$, it is seen that 
	the cone vertex will lie outside the ellipsoid whenever   
	\begin{equation}
	p_2 \geq \frac{p_1 \sin^2\beta_1}{2(1-\cos\beta_1\cos\gamma)}\,,\label{cv}
	\end{equation}
	(if $p_1<0$, the inequality should be inverted). The straight lines delimiting the cone end at the ellipsoid points $(x,y,z)$ 
	satisfying $\frac{x(x-x_v)+y(y-y_v)}{b^2}+\frac{(z-z_c)(z-z_v)}{a^2}=0$.  We also note that the entanglement of  (\ref{rh2}) 
	is in this case driven just by $|\Psi_1\rangle$,  with the partial transpose becoming non-positive 
	just for $p_1\sin\beta_1>p_0/d_A$, i.e., $p_1>(1-p_2)/(1+d_A\sin\beta_1)$. 
	
\begin{figure}
		\hspace*{0.cm}\includegraphics[width=.48\textwidth]{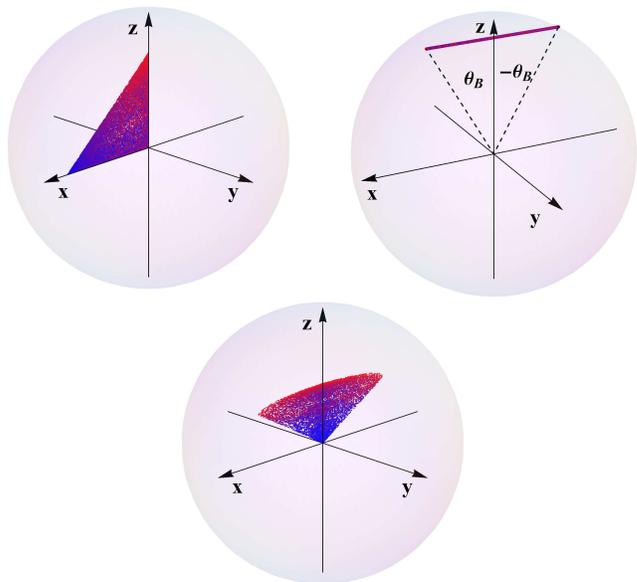}
		\vspace*{0cm}
				\caption{The set of post-measurement qubit states after random measurements on the qudit, for the mixture (\ref{rh2}) with  both $|\Psi_1\rangle$ and  $|\Psi_2\rangle$ separable ($\beta_1=\beta_2=0$).  For $p_1+p_2<1$ and $d_A\geq 3$, 
			it is  a two-dimensional  triangle (if $\gamma\in(0,
			\pi)$ in (\ref{cone})),  as shown on the top left panel for $p_1=0.3$,  $p_2=0.2$, $\gamma=\pi/2$, $\eta=0$ and  $d_A=4$. When $p_1+p_2=1$, the triangle reduces to the segment joining the pure states Bloch vectors, as shown schematically on the top right panel. 
			Such needle-like shape holds for any  mixture of two arbitrary pure separable states like   Eqs.\ (\ref{rhasb1})--(\ref{rhasb2}), 
			and any $d_A\geq 2$. On the other hand, for $p_1+p_2<1$, $d_A\geq 3$ and $|\Psi_{1,2}\rangle$  separable but with non-orthogonal supports at the qudit side, the set  becomes again triangle-like (and flat) but with a  rounded outer border, as shown in the bottom panel  for $p_1=p_2$ (see text).} 
		\label{f5}
	\end{figure}

	{\it Triangles and segments.}
	If $|\Psi_1\rangle$ also becomes separable ($\beta_1=0$),  the first ellipsoid  reduces to a segment, linking the origin with $\bm{r}_{1}=p_1(0,0,1)/(p_1+p_0/d_A)$. The ensuing convex hull of both segments leads to a 
	{\it two-dimensional triangle} if they  are non-collinear, i.e., $\gamma\neq 0$ in (\ref{rv}), with vertices at the origin,  $\bm{r}_1$ and $\bm{r}_2=\bm{r}_v$,  as seen in Fig.\ \ref{f5} (top left). 
	This result holds whenever the maximally mixed state has non-zero weight, i.e., $p_1+p_2<1$ and $d_A\geq 3$. 
	
	On the other hand, if $p_1+p_2=1$, the previous segments reduce to the points $\bm{r}_1$ and $\bm{r}_2$  
	on the Bloch sphere and their convex hull becomes just the {\it segment} between them, leading to a {\it needle}-type state  
	like that of the top right panel in Fig.\ \ref{f5}. The final set does not depend on the ratio $p_2/p_1$ as long as both probabilities are non-zero, 
	and holds in this case for any  $d_A\geq 2$. 

	\subsection{Mixture of two separable pure states} 
	\label{II.D}
	Needle shapes actually emerge from {\it any}  mixture of two pure product states, i.e., 
	\begin{eqnarray}
	\rho_{AB}&=&
	p_1|\Psi_1\rangle\langle\Psi_1|+p_2|\Psi_2\rangle\langle\Psi_2|\,,\label{rhasb1}\\
	|\Psi_{i}\rangle&=&|\psi^A_{i}\rangle|\psi^B_{i}\rangle\,,\;\;i=1,2\,,\nonumber
	\end{eqnarray}
	where {\it both} local states $|\psi^{A}_{i}\rangle$, $|\psi^{B}_{i}\rangle$, $i=1,2$, are now completely arbitrary (and $p_1+p_2=1$). The mixture (\ref{rhasb1}) can be conveniently rewritten as 
	\begin{eqnarray}
	\rho_{AB}&=&p_+|\theta_A\theta_B\rangle	\langle \theta_A\theta_B|+
	p_-|\text{-}\theta_A\text{-}\theta_B\rangle\langle 
	\text{-}\theta_A\text{-}\theta_B|\,, \label{rhasb2}
	\end{eqnarray}
	 where $p_{+(-)}=p_{1(2)}$ and $|\pm\theta_{S}\rangle=e^{\mp i\phi_S/2}|\psi_{1,2}^S\rangle$ 
	 for $S=A,B$,  with $\phi_S$ and $\theta_S\in[0,\pi/2]$ determined by 
	\begin{equation}\langle\psi_2^S|\psi_1^S\rangle=e^{i\phi_S}
	\cos\theta_{S}\,,\;\;S=A,B\,,\end{equation}
	such that $\langle-\theta_S|\theta_S\rangle=\cos\theta_S$. 
	In this way, $|\pm\theta_S\rangle=
	\cos\tfrac{\theta_S}{2}|0_S\rangle\pm\sin\tfrac{\theta_S}{2}|1_S\rangle$ can be seen  as  qubit states  
	rotated an angle $\pm\theta_{S}$ from states $|0_{S}\rangle$ around the $y$ axis,  with  
	\begin{equation}
	|0_S\rangle=\frac{|\theta_S\rangle+|-\theta_S\rangle}{2\cos\tfrac{\theta_S}{2}},\;\;|1_S\rangle=
	\frac{|\theta_S\rangle-|-\theta_S\rangle}{2\sin\tfrac{\theta_S}{2}}\,,
	\label{01A}\end{equation}
	orthonormal states.
	
	A general pure state $|\bm{k}_A\rangle$ of qudit $A$  can now be written as in Eq.\ (\ref{kq}), 
	with $|\bm{k}^{\parallel}_A\rangle$ of the form (\ref{pw}). 
    	The ensuing conditional state of $B$, 
	\begin{equation}
	\rho_{B/\bm{k}}=
	p'_+(\bm{k})|\theta_B\rangle\langle\theta_B|+ p'_-(\bm{k})|\text{-}\theta_B\rangle\langle\text{-}\theta_B|\,.\label{rbkt}
	\end{equation}
	has the same form as the original state $\rho_B={\rm Tr}_A \rho_{AB}=p_+|\theta_B\rangle\langle\theta_B|
	+p_-|\text{-}\theta_B\rangle\langle\text{-}\theta_B|$, but with modified weights 
	\begin{equation}
	p'_{\pm}(\bm{k})=\frac{p_{\pm}|\langle\bm{k}^\parallel_A|
		\pm\theta_A\rangle|^2}
	{p_+|\langle\bm{k}^\parallel_A|\theta_A\rangle|^2+p_-|\langle\bm{k}^\parallel_A|\text{-}\theta_A\rangle|^2}\,,
	\end{equation}
	which  cover  {\it all} values between $0$ and $1$ as $|\bm{k}^\parallel_A\rangle$ is varied (since  
	$p'_{\pm}(\bm{k})=0$,  when $|\bm{k}^\parallel_A\rangle$ is orthogonal to $|\pm\theta_A\rangle$).  Consequently, 
	the set of postmeasurement Bloch vectors $\bm{r}_{B/\bm{k}}=\sum_{\nu=\pm}p'_{\pm}(\bm{k})\bm{r}_{B\pm}$, with $\bm{r}_{B\pm}=\langle \pm\theta_B|\bm{\sigma}|\pm\theta_B\rangle=(\pm\sin\theta_B,0,\cos\theta_B)$,  is always the full segment joining the 
	points  $\bm{r}_{B\pm}$ located on the Bloch sphere surface,  
	as shown on the top right panel of Fig.\ \ref{f5}. 
	
	This result holds for any qudit dimension $d_A\geq 2$, and is then similar to that for a two-qubit system in a similar state (the needle is in fact a limit case of an ellipsoid),  since the local support at $A$ of the state (\ref{rhasb1}) is two-dimensional. Differences with the two-qubit  case arise only when the  support  of $\rho_{AB}$  involves a qudit subspace of higher dimension, as was shown in the examples of Figs.\ \ref{f1}, \ref{f3}, \ref{f4} and \ref{f5} (top left).   
	
	For example, if we now mix the state (\ref{rhasb1})--(\ref{rhasb2}) with the maximally mixed state, as in Eq.\ (\ref{rh2}),  
	$\rho_{B/\bm{k}}$ will be a mixture of the state (\ref{rbkt}) with the maximally mixed state $\mathbbm{1}_{B}/2$. Using (\ref{kq})--(\ref{pw}), the ensuing conditional Bloch vector becomes  $\bm{r}_{B/\bm{k}}= q  \sum_{\nu=\pm}\tilde{p}_{\pm}(\bm{k})\bm{r}_{B\pm}$, with $\tilde{p}_{\pm}(\bm{k})=p'_{\pm}(\bm{k})/(q+p_0/[d_A\sum_{\nu=\pm}	|\langle\bm{k}_A^{\parallel}|\nu\theta_A\rangle|^2])$ 	and $p_0=1-p_+-p_-$. In the two-qubit case $d_A=2$,  $q=1$ and the set of conditional vectors will form a flat filled ellipse (``pancake shape'', limit case of an ellipsoid surface)  in the $xz$ plane. 
	However, for $d_A\geq 3$ qudit states orthogonal to both  $|\pm\theta_A\rangle$ exist and hence $q\in[0,1]$, implying that such set will become the convex mixture of a similar flat filled ellipse with the origin. It will  lead to a flat shape like that shown at the bottom of Fig.\ \ref{f5}, i.e.\ triangle-like but with a rounded upper border, obtained for $p_+=p_-=0.3$,  $d_A=4$ and $\theta_A=\pi/4$, $\theta_B=\pi/6$.

	\section{Minimum Conditional entropy}
	\label{III}
	
	\subsection{Measurement determined conditional entropy}
	\label{III.a}
	
	The concept of a measurement determined  quantum conditional entropy of a bipartite system was originally introduced  in connection with the quantum discord  \cite{OZ.01,HV.01,Modi.12,ABC.16,Lec.17,Bera.18}.  It is a measure of the average conditional mixedness of the unmeasured subsystem $B$ after a local measurement at $A$, and its minimum over all local measurements determines the quantum discord. Such minimization is in general a difficult problem, shown to be  NP complete \cite{Huang.14}.  The concept was later extended to generalized entropic forms \cite{GR.14,GRb.14}, which can enable a simpler evaluation and an analytic determination of the minimum in some cases.  Here we will discuss the generalized conditional entropy in the states considered in previous section, providing analytic results for its minimizing measurement and its geometric picture. Such  measurement is also interesting in itself, since it maximizes the average amount of information on $B$ that can be gained  through  measurements at $A$.
	
	Given a local measurement at $A$ determined by measurement operators $M_j=M_j^A\otimes \mathbb{1}_{B}$, 
	satisfying $\sum_j M_j^\dagger M_j=\mathbb{1}_{AB}$, 
	the generalized measurement dependent conditional entropy $S_f(B|A_{M})$ is 
	defined as \cite{GRb.14,GR.14}
	\begin{equation}
	S_f(B|A_{M})=\sum_j p_{j}S_f(\rho_{B/j})\,,
	\label{Scond}
	\end{equation}
	where $p_j={\rm Tr}\,[\rho_{AB}M_j^\dagger M_j]$ is the probability of outcome $j$ and $\rho_{B/j}=p_j^{-1}{\rm Tr}_A\,[\rho_{AB}M_j^\dagger M_j]$  
	the conditional state of $B$ after this outcome. Here 
	\begin{equation}
	S_f(\rho) = \text{Tr} f(\rho),\label{gS}
	\end{equation}
	is a trace form entropy \cite{W.78,CR.02}, 
	with $f: [0,1] \rightarrow \mathbb{R}$ a smooth strictly concave function satisfying $f(0) = f(1) = 0$ 
	(implying  $S_f(\rho)$ concave and $S_f(\rho)\geq 0$, with $S_f(\rho)=0$ iff $\rho$ is a pure state). 
	For $f(p) = - p \log_2 p$, (\ref{gS}) becomes the von Neumann entropy $S(\rho)=-{\rm Tr}\rho\log_2\rho$, 
	while for $f(p) = 2p (1-p)$, it becomes the linear entropy
	\begin{equation}
	S_2(\rho)=2(1-{\rm Tr}\,\rho^2), \label{S2} 
	\end{equation}
	also known as quadratic entropy and coincident with the $q=2$ Tsallis entropy \cite{Tsa.88},   obtained for $f(p)\propto\frac{p-p^q}{q-1}$.  
	
	Since $\rho_B=\sum_j p_j\rho_{B/j}$,  concavity of $S_f$ directly implies $S_f(B)\equiv S_f(\rho_B)\geq \sum_j p_j S_f(
	\rho_{B/j})=S_f(A|B_M)$, 
		with equality iff $\rho_{B/j}=\rho_B$ $\forall$ $j$ with $p_j>0$. Thus, the generalized conditional entropy is never greater than the corresponding marginal entropy. 
	
	Its minimum over all local measurements at $A$, 
	\begin{equation}S_f(B|A)=\mathop{\rm Min}_{M}\,S_f(B|A_{M})\,,\label{Smin}\end{equation}
	depends just on $\rho_{AB}$, with $\Delta S_f(B|A)=S_f(B)-S_f(B|A)$ a nonnegative quantity  
	that measures the maximum average conditional {\it information gain} about $B$, as measured by $S_f$, that can be obtained through a measurement at $A$. 
	$\Delta S_f(B|A)$ vanishes iff $\rho_{AB}$ is a product state. The minimum (\ref{Smin})  also represents  the generalized entanglement 
	of formation \cite{BD.96}  $E_f(B,C)$  of system $B,C$, where $C$ is a third system purifying the whole system \cite{KW.02,GRb.14}.  
	In the case of the von Neumann entropy, the minimum $S(B|A)$ determines  the  quantum discord  through 
	\cite{OZ.01,HV.01} $D(B|A)=S(B|A)-\tilde{S}(B|A)$, where  $\tilde{S}(B|A)=S(\rho_{AB})-S(\rho_A)$ is 
	the standard quantum  conditional entropy \cite{W.78}. While the 
	latter  is negative in any  pure entangled state, the minimum (\ref{Smin})  is obviously always  nonnegative, vanishing in any pure state \cite{GR.14}.  
	
	\subsection{Minimizing measurement}
	
	The minimum (\ref{Smin}) can be always reached for measurements based on rank-$1$  operators $M_j=\sqrt{r_j}\Pi^A_{\bm{k}_j}$ \cite{Modi.12,ABC.16},	of the type considered in section \ref{II},  a  result which holds for any $S_f$ \cite{GR.14,GRb.14}. 
	It is then sufficient to consider 
	\begin{equation} 
	S_f(B|A_{M})=\sum_j r_j p_{\bm{k}_j} S_f(\rho_{B/\bm{k}_j}) \,,
	\label{Scondk}
	\end{equation}
	with $\rho_{B/\bm{k}_j}$ given by Eq.\ (\ref{cqu}) and projectors $\Pi_{\bm{k}}^A$ within the local  support at $A$ of $\rho_{AB}$, as discussed below Eq.\ (\ref{kq}).
	
	For a mixture of a single pure state $|\Psi\rangle$ with the maximally mixed state, Eq.\ (\ref{rh1}),  
	the minimum (\ref{Smin}) is reached, for {\it any} $S_f$,  {\it for a projective measurement on the  local Schmidt basis}  
	$\{|k_A\rangle\}$ determined by the state $|\Psi\rangle$   \cite{GRb.14}. In the appendix A it is shown that this result 
	{\it can be extended to any mixture of the form (\ref{rh2})}, where the local supports at $A$ of the states $|\Psi_i\rangle$ 
	are orthogonal and hence compatible with a unique local Schmidt basis: 
	\begin{eqnarray}\rho_{AB}&=&\sum_{i=1}^n p_i |\Psi_i\rangle\langle\Psi_i|+p_0\frac{\mathbbm{1}_{AB}}{d_A d_B}\,,\label{rabg}\\
	|\Psi_i\rangle&=&\sum_k \sqrt{q_{ik}}|k_A\rangle|k_{iB}\rangle\,,
	\label{psii}\end{eqnarray}
	where $p_0=1-\sum_{i=1}^n p_i\geq 0$, $p_i\geq0$ and $q_{ik}=\delta_{i,i_k}q_{k}$, $q_k>0$, such that for 
	each $k$ there is at most a single state  $|\Psi_{i_k}\rangle$ with finite overlap with $|k_A\rangle$.  
	Hence, for a measurement in the basis $\{|k_A\rangle\}$, 
	\begin{equation}\rho_{B/k}=\frac{1}{p_k^A}\left(p_{i_k}q_k
	|k_{i_k B}\rangle\langle k_{i_k B}|+p_0\frac{\mathbbm{1}_B}{d_A d_B}\right)\,,\label{rbkk}\end{equation}
	where $p_k^A=p_{i_k}q_k+p_0/d_A$. The ensuing minimum conditional entropy is then   
	\begin{equation}
	S_f(B|A)=\sum_k p_k^A[f(\tfrac{d_A d_Bp_{i_k}q_k+p_0}{d_A d_B p_k^A})+
	(d_B-1)f(\tfrac{p_0}{d_Ad_B p_k^A})]\,.
	\label{sfmin}\end{equation}
	In the von Neumann case, this expression enables a direct evaluation of the quantum discord $D(B|A)$. 
	
	Consequently, for the states of secs.\ \ref{II.b}--\ref{II.c}, the post-measurement states of the qubit 
	determined by the minimizing measurement at the qudit have  a clear geometric picture:  In the  state 
	(\ref{rh1})--(\ref{psi}), the minimum is obtained for a projective measurement  in a basis containing  
	the states $|0_A\rangle$ and 
	$|1_A\rangle$, i.e., $\alpha=0$ and $\pi$ in (\ref{pw}), with $q=1$.  The associated conditional qubit 
	Bloch vectors lie at  the  {\it ellipsoid extrema} along the major ($z$) axis. Similarly, for the states (\ref{rh2}) 
	the minimizing measurement basis should contain in addition the states $|2_A\rangle$ and  $|3_A\rangle$, 
	and the ensuing conditional qubit vectors lie at the ellipsoids major axes extrema. 
	For  $\beta_2=0$ (``ice-cream'' shapes), this leads to the   ellipsoid extrema and the cone vertex, 
	i.e., $q'=1$, $q=1$, $\alpha_1=0,\pi$, and $q=0$, $\alpha_2=0$ in (\ref{kag})--(\ref{kia}), while for a triangle shape, to the triangle vertices. 
	
	\subsection{The quadratic conditional entropy\label{III.C} }
	For more general states $\rho_{AB}$,  the problem of determining the minimizing measurement is in general hard \cite{Huang.14}. It is then convenient to consider the quadratic conditional entropy derived from (\ref{S2})  \cite{GR.14}, which is  determined by the state purity and hence does not require the knowledge  of its eigenvalues,  and which can in principle be accessed experimentally without the need of a full state tomography \cite{NK.12}. 
	First, by means of  Eqs. (\ref{ort}) and (\ref{rhos}), the quadratic marginal entropy can be evaluated explicitly as 
	\begin{equation}
	S_2(\rho_S) =\frac{2}{d_S}(d_S-1-|\bm{r}_S|^2), \;\;S=A,B\,, \label{S2k}
	\end{equation}
	which shows that $|\bm{r}_S|^2 \leq d_S-1$, with equality  iff $\rho_S$ is pure. 
	The corresponding conditional entropy (\ref{Scondk}) can also be explicitly determined using Eqs.\ (\ref{cqu})--(\ref{rac}) \cite{GR.14}: 
	\begin{equation}S_2(B|A_M)=
	S_2(\rho_B)-\Delta S_2(B|A_M)\,,\label{S2cond} 
	\end{equation}
	where 
	\begin{equation}
	\Delta S_2(B|A_M)=\frac{1}{d_B}\sum_j r_j\frac{|C^T\bm{k}_j|^2}{1+\bm{r}_A\cdot\bm{k}_j}\label{Del1}
	\end{equation}
	is a nonnegative quantity representing an information gain. Here $C$ is the correlation tensor (\ref{cten}) 
	and $\bm{r}_A=\langle\bm{\sigma}_A\rangle$. 
	
	In particular, if the local support at $A$ of state $\rho_{AB}$ involves just {\it two}  pure states $|0_A\rangle$, $|1_A\rangle$, 
	we may directly consider projectors within this subspace ${\cal S}$ and use effective Pauli operators at system $A$. 
	For a standard projective measurement based on the orthogonal states $|\pm \bm{k}_A\rangle$, 
	with $\langle\pm\bm{k}_A|\bm{\sigma}|\pm\bm{k}_A\rangle=\pm\bm{k}$,  Eq.\ (\ref{Del1}) reduces to \cite{GR.14}
	\begin{equation}\Delta S_2(B|\bm{k}_A)=\frac{2}{d_B}\frac{|C^T\bm{k}|^2}{1-(\bm{r}_A
		\cdot\bm{k})^2}=\frac{2}{d_B}\frac{\bm{k}^TCC^T\bm{k}}
	{\bm{k}^TN_A\bm{k}}\,,\label{S2x}
	\end{equation}
	where $N_A=\mathbbm{1}-\bm{r}_A\bm{r}_A^T$. 
	Maximization of (\ref{S2x}) (equivalent to minimization of (\ref{S2cond})) over these measurements is then 
	achieved by solving the weighted eigenvalue problem 
	\begin{equation}
	CC^T\bm{k}=\lambda N_A\bm{k}\,,\label{eig2}
	\end{equation}
	and selecting the largest
	eigenvalue $\lambda_{\rm max}$, 
	with the optimizing measurement determined by the corresponding eigenvector $\bm{k}$ (i.e., it is an effective 
	spin $1/2$ measurement in ${\cal S}$ along direction $\bm{k}$). This leads to
	\begin{equation}
	\mathop{\rm Min}_{\bm{k}}S_2(B|{\bm{k}_A})=S_2(\rho_B)-\frac{2}{d_B}\lambda_{\rm max}\,.
	\label{S2min}
	\end{equation}
	More general POVM measurements based on an arbitrary  set $\{r_j \Pi^A_{\bm{k}_j}\}$ do not improve  previous minimum \cite{GR.14}.  
	
	Hence, for these states the linear entropy allows a direct analytic evaluation of the associated minimum conditional entropy and its minimizing measurement. As a check, for a two-qubit state $\rho_{AB}=p|\Psi\rangle\langle\Psi|+(1-p)\mathbb{1}/4$, with $|\Psi\rangle$ 
	of the form (\ref{psi}), $\bm{r}_{A(B)}=(0,0,p\cos\beta)$ and  $C_{\mu\nu}=\delta_{\mu \nu} C_{\mu\mu}$, with 
	$C_{xx}=-C_{yy}=p\sin\beta$, $C_{zz}=p(1-p\cos^2\beta)$. It is then verified that the largest eigenvalue of 
	Eq.\ (\ref{eig2}) is $\lambda_z=C_{zz}^2/(1-p^2\cos^2\beta)$ ($\lambda_z>\lambda_x=\lambda_y=p^2\sin^2\beta$),  
	associated to eigenvector $\bm{k}_z=(0,0,1)$, implying measurement in the Schmidt basis 
	$\{|0_A\rangle, |1_A\rangle\}$. Eq.\ (\ref{S2min}) then coincides with (\ref{sfmin}) for $f(p)=2p(1-p)$.  
	
	\subsection{The case of rank-$2$ separable states} 
	Previous expressions enable to determine the minimum conditional entropy in the state (\ref{rhasb2}) and the associated 
	minimizing measurement. Of course, if $\theta_A=\pi/2$, states 	$|\pm\theta_A\rangle$ are orthogonal and a projective measurement 
	on this basis, i.e., a spin measurement along the $x$ axis in the qubit picture, 
	provides the minimum ($S_f(A|B)=0$). 
	
	For general $\theta$, we obtain $C_{\mu\nu}=\delta_{\mu\nu}\delta_{\mu,x}C_{xx}$, with  
	$C_{xx}=4p_+p_-\sin\theta_A\sin\theta_B$, while for $S=A,B$,  $\bm{r}_{S}=((p_+-p_-)\sin\theta_{S},0,\cos\theta_{S})$. 
	 The matrix $N_A$ in (\ref{eig2}) is then non diagonal  if $p_+\neq p_-$ and $\theta_A\in(0,\pi/2)$, 
	and Eq.\ (\ref{eig2}) leads  to  $\bm{k}=(\sin\phi,0,\cos\phi)$, with 
	\begin{equation}
	\tan\phi=\frac{\tan\theta_A}{p_+-p_-}\,,
	\label{result}
	\end{equation}
	if  $p_+\geq p_-$ (i.e.\ $p_+\geq 1/2)$. This entails a spin-like measurement at the effective qubit along a direction $\bm{k}$ 
	forming an angle $\phi$ with the $z$ axis, such that 
	\begin{equation}
	|\bm{k}_A\rangle=|\phi_A\rangle,\;\;|-\bm{k}_A\rangle=
	|(\phi+\pi)_A\rangle\,.\label{kaa}
	\end{equation} 
	
	The meaning of the minimizing angle (\ref{result}) is that the ensuing entropies $S_2(\rho_{B/\pm\bm{k}})$ are {\it equal}, 
	i.e., the vectors $\bm{r}_{B/\pm\bm{k}}$ have both the same length:   
	\begin{equation}
	\bm{r}_{B/\pm\bm{k}}=(\pm\sin\theta_B\sqrt{(p_+-p_-)^2\cos^2\theta_A+\sin^2\theta_A},0,\cos\theta_B)
	\end{equation}
	These vectors are not  at the edges $(\pm \sin\theta_B,0,\cos\theta_B)$ of the segment (except when the states 
	$|\pm\theta_A\rangle$ are orthogonal, i.e. $\theta_A=\pi/2$),  but rather at inner {\it symmetric} points with respect to the $z$ axis.
	The ensuing minimum conditional entropy (\ref{S2min}) is just 
	\begin{equation} 
	S_2(B|A)=1-|\bm{r}_{B/\pm \bm{k}}|^2=4p_+p_-\cos^2\theta_A\sin^2\theta_B\,.\label{Scondmin}
	\end{equation}
	It vanishes in the trivial cases $\theta_A=\pi/2$ or $\theta_B=0$. 
	
	In the equally weighted case $p_+=p_-=1/2$,  Eq.\ (\ref{result}) leads to $\phi=\pi/2$ {\it for any $\theta_A$}, 
	i.e.\ to a spin measurement along the $x$ axis (states $|\pm\bm{k}_A\rangle=\frac{|0_A\rangle\pm|1_A\rangle}{\sqrt{2}}$), 
	in agreement with the fact that $N_A$ becomes diagonal and the  only non-zero correlation is $C_{xx}$. 
	The solution (\ref{result}) can actually be also  obtained in this way, by considering (\ref{rhasb2}) as an equally weighted mixture of unnormalized states 
	$\sqrt{p_{\pm}}|\!\pm\theta_A,\!\pm\theta_B\rangle$. The normalized (but non-orthogonal) states $|0'_A\rangle$, $|1'_A\rangle$  associated with the latter are 
	$^{|0'_A\rangle}_{|1'_A\rangle}=\frac{\sqrt{p_+}|\theta_A\rangle\pm\sqrt{p_-}|\text{-}\theta_A\rangle}{\sqrt{1\pm 2\sqrt{p_+p_-}\cos\theta_A}}$,  
	and the ensuing normalized orthogonal states along $x'$, 
	$|\pm \bm{k}_A\rangle=\frac{|0'_A\rangle\pm|1'_A\rangle}{\sqrt{2(1\pm\langle 1'_A|0'_A\rangle)}}$, 
	are precisely the states (\ref{kaa}) determined by  Eq.\ (\ref{result}). 
	
	A remarkable feature is that Eq.\  (\ref{result}) determines the minimizing measurement for {\it any conditional 
	entropy} $S_f(B|A_M)$ for which the entropy $S_f(\rho)=f(p)+f(1-p)$ of a single qubit state $\rho$ is a {\it convex} increasing function of $\sqrt{S_2(\rho)}=2\sqrt{p(1-p)}$ 
	($S_f(\rho)=F\left(\sqrt{S_2(\rho)}\right)$, with $F(x)=\sum_{\nu=\pm}f\left(\frac{1+\nu\sqrt{1-x^2}}{2}\right)$).   The reason  is  that for these states the  system $C$ purifying 
	the whole system is also a qubit, 
	and hence,  for these entropies the entanglement of formation 
	$E_f(B,C)=S_f(B|A)$ is determined  by the concurrence \cite{WW.97}, which is just $\sqrt{E_2(B,C)}=\sqrt{S_2(B|A)}$.  
	Thus,  $S_f(B|A)=E_f(B,C)=F(\sqrt{E_2(B,C)})=F(\sqrt{S_2(B|A)})$. 
	And since the $S_2$ minimizing measurement leads to coincident post-measurement entropies $S_f(\rho_{B/\pm\bm{k}})=F(\sqrt{S_2(B|A)}$,  
	it also minimizes $S_f(A|B_M)$ for such entropies. Convexity of $F$  holds, in particular, for the Tsallis entropies with  $\frac{5-\sqrt{13}}{2}\leq q\leq \frac{5+\sqrt{13}}{2}$ \cite{CRCb.10}, 
	{\it including} the von Neumann entropy 
	(recovered for $q\rightarrow 1$). 
	Hence, present results also enable a direct evaluation of the quantum discord for these states, and are in agreement with those of \cite{Shi.11}.

	\subsection{Mixture of aligned spin-$s$ states}
\begin{figure}
		\begin{center}
			\includegraphics[width=0.45\textwidth]{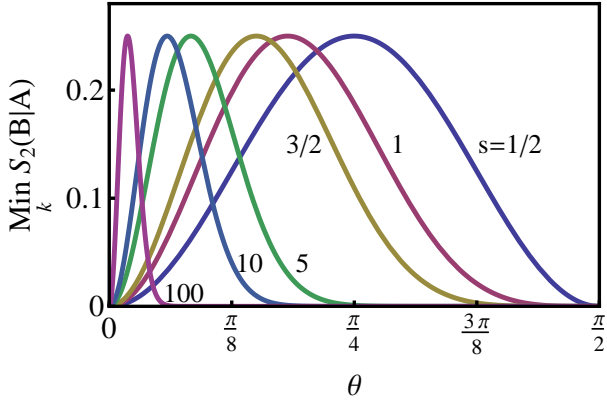}
		\end{center}
		\vspace*{-0.42cm}
		
		\caption{The minimum conditional entropy (\ref{QCs}) as a function of $\theta$ for the 
			indicated values of the spin $s$ in the state (\ref{thes}).}
		\label{f6}
	\end{figure}

	As application of previous result, we finally consider the case of two actual spins $s$  
	in a mixture of two maximally aligned states $|\!\!\nearrow\nearrow\rangle=|\theta_s\theta_s\rangle$ and  
	$|\!\!\nwarrow\nwarrow\rangle=|\text{-}\theta_s\text{-}\theta_s\rangle$, in directions forming angles  $\pm\theta$  
	with the $z$ axis. These states arise, for instance, as exact reduced pair states in  the 
	ground state of  $XY$ and $XYZ$ spin chains in an applied transverse field along $z$, 
	in the immediate vicinity of the factorizing field \cite{RCM.08,CRM.10}. 
	Their joint state takes the form (\ref{rhasb2}), i.e., 
	\begin{equation}
	\rho_{AB}=p_+|\theta_s\theta_s\rangle\langle\theta_s\theta_s|+p_-|\text{-}\theta_s\text{-}\theta_s\rangle
	\langle\text{-}\theta_s\text{-}\theta_s|\,,\label{thes}
	\end{equation}
	with $p_\pm=1/2$ and 	$|\pm\theta_s\rangle= e^{\mp \imath\theta S_{y}} |m=s\rangle$ given by 
	\begin{equation}
	|\pm\theta_s\rangle
	=\sum_{m=-s}^{s}\sqrt{
		(^{2s}_{m+s})} \cos^{s+m} (\tfrac{\theta}{2})\sin^{s-m}(\tfrac{\pm\theta}{2})
	|m\rangle\label{ths}\,.
	\end{equation}
	Since $\langle-\theta_s|\theta_s\rangle = \cos ^{2s} \theta_s$, 
	these states correspond to an effective qubit angle $\theta_A=\theta_B$ in (\ref{rhasb2}), with  $\cos\theta_A=\cos^{2s}\theta$. Hence, according to Eq.\ (\ref{Scondmin}), the minimum $S_2$ conditional entropy is  
	\begin{equation}
	S_2(B|A) = 4 p_+ p_- \cos ^{4s} \theta (1-\cos^{4s} \theta)\,.\label{QCs}
	\end{equation}
	In Fig.\ \ref{f6} we depict $S_2(B|A)$ vs.\ $\theta$ for different values of the spin $s$ in the equally probable case  $p_\pm=1/2$. 
	Its maximum is $s$-independent but is  reached at $\theta=\theta_m=\arccos(2^{-\frac{1}{4s}})$, 
	which vanishes for large $s$ ($\theta_m\approx\sqrt{\frac{\ln 2}{2s}}$ for $s\gg 1$). The minimum conditional entropy $S_2(B|A)$ 
	becomes then very small for $\theta>\theta_m$ and large $s$, as states  $|\pm\theta_s\rangle$ become  almost orthogonal. 
	
	\begin{figure}
		\includegraphics[width=0.45\textwidth]{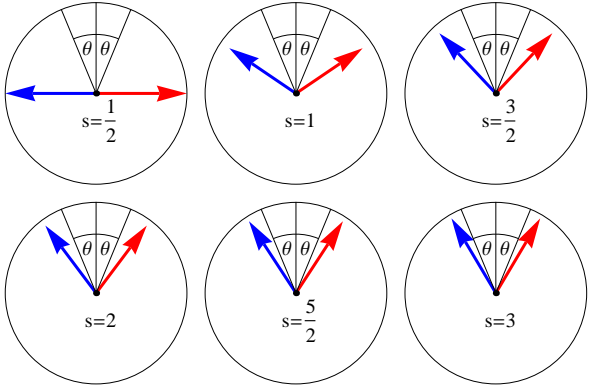}
		\vspace*{-0.25cm}
		
		\caption{The scaled averages (\ref{Sav}) of the spin operator $\bm{S}$ in the orthogonal  states $|\pm\bm{k}_A\rangle$ which 
			minimize the $S_2$ conditional entropy,  for the joint state (\ref{thes}). As the spin $s$ increases they 
			approach the $\pm\theta$ directions. } 
		\label{f7}
	\end{figure}
	
	On the other hand, the local measurement minimizing the conditional entropy is that in an orthogonal basis of states 
	containing the states (\ref{kaa}), where 
	$|\phi_A\rangle=\cos\tfrac{\phi}{2}|0_A\rangle+\sin\tfrac{\phi}{2}|1_A\rangle$ with (Eq.\ (\ref{01A}))  
	$|^{0_A}_{1_A}\rangle=\frac{|\theta_s\rangle\pm|-\theta_s\rangle}{\sqrt{2(1\pm\cos^{2s}\theta)}}$. For $s\geq 1$,  
	such measurement {\it is not a spin measurement}, in the sense that it does not correspond to the 
	measurement of the spin at a certain direction $\bm{n}$: In the latter, the spin has collinear integer values 
	$\langle m_{\bm{n}}|\bm{S}|m_{\bm{n}}\rangle=m\bm{n}$, with $m=-s,\ldots,s$, 
	whereas in the minimizing states (\ref{kaa}), 
	it takes non-parallel and non-integer average values, as occurs for general projective measurements (see \cite{PRA.12}). For $p_\pm=1/2$ and $\phi=\pi/2$  we obtain  
	\begin{equation}\langle \pm\bm{k}_A| \bm{S} |\pm\bm{k}_A\rangle/s={\textstyle\left(\pm\frac{\sin\theta}
	{\sqrt{1-\cos^{4s}\theta}},0,\frac{\cos\theta (1-\cos
		^{4s-2}\theta)}{1-\cos^{4s}\theta}\right)}\,.
	\label{Sav}\end{equation}
		As seen in Fig.\ \ref{f7}, while for $s=1/2$ the vectors  $\langle \bm{s}\rangle_{\pm}=\langle\pm\bm{k}_A|\bm{S}|\pm\bm{k}_A\rangle/s$ 
	point along the $x$ axis, indicating a spin measurement along the $x$  axis, for $s\geq 1$ they are noncollinear and approach, 
	for large $s$, the $\pm\theta$ directions, i.e.,  $\langle \bm{s}\rangle_{\pm}\approx\langle \pm\theta_s|\bm{S}|\pm\theta_s\rangle/s=
	(\pm\sin\theta,0,\cos\theta)$, coinciding with  the latter for $s\rightarrow\infty$. This result is to be expected as in this limit  
	the states $|\pm \theta_s\rangle$ become orthogonal. In the case of spin $s=1$, the averages (\ref{Sav}) imply that the minimizing 
	measurement is an Y-type projective measurement,  following the terminology of \cite{PRA.12}, based on the states $|\pm\bm{k}_A\rangle$ 
	and a third state orthogonal to the latter.

	\section{Conclusions}
	
	We have first shown that in correlated mixed states of qudit-qubit systems, the set of all conditional qubit states after a general local measurement at the 
	qudit based on rank-$1$ projectors, may exhibit geometries which are more complex than a single ellipsoid. While a single solid ellipsoid, 
	with the origin as one of its foci, is obtained for a state which is the  mixture of a pure entangled state 
	with the maximally mixed state, for more general mixtures of the form (\ref{rh2}) such set becomes  the convex hull of different solid ellipsoids, 
	leading to  shapes like that of Fig.\ \ref{f3}. These shapes may become cone-like or flat triangle-like  when one or more of the pure states 
	of the mixture are separable, as shown in Figs.\ \ref{f4}--\ref{f5}. 
	
	We have also analyzed the corresponding measurement dependent conditional entropy and  its minimizing measurement for the previous states. 
	For a mixture of a single pure state with the maximally mixed state, such  measurement  is that on the pure state Schmidt 
	basis and is universal, in the sense of minimizing  {\it any} entropy of the form (\ref{gS}).  We have shown that this result 
	can be extended to any mixture (\ref{rh2})  where the local supports  of the states $|\Psi_i\rangle$ at the qudit side  are orthogonal, leading to a clear geometric 
	interpretation of the minimizing measurement in the set of post-measurement qubit states. These minimizing measurements maximize the average conditional information gain about $B$, and enable to determine the quantum discord.  We also examined the case 
	of rank-$2$ separable states (\ref{rhasb1})--(\ref{rhasb2}), determining the minimizing measurement analytically for 
	a wide class of entropies through the linear entropy. As application, the minimizing measurement for a mixture of 
	maximally aligned two-spin states was determined for general spin $s$, and shown not to correspond to a standard spin measurement 
	for any spin $s\geq 1$. The experimental verification of these results through optical means is currently  under investigation.

	\acknowledgments
	Authors acknowledge support from CONICET (MB, NC, NG) and  CIC (RR)  of Argentina.
	\appendix*
	\section{}
	We prove here that the measurement at the qudit $A$ minimizing the general conditional entropy (\ref{Scond}) for states of the form 
	\eqref{rabg}--\eqref{psii}, where the states  $|\Psi_i\rangle$ have orthogonal local supports at $A$, is  on the local Schmidt basis.

	{\it Proof:} For a local measurement  based on rank-$1$  operators  $\sqrt{r_j}|j_A\rangle\langle j_A|$, 
	with $\sum_j r_j|j_A\rangle\langle j_A|=\mathbbm{1}_{d_A}$, we have  
	$\rho_{AB/j}=\Pi^A_j\rho_{AB}\Pi_j^A/p_j^A$, where $\Pi_j^A=|j_A\rangle\langle j_A|\otimes \mathbbm{1}_{d_B}$ and 
	\begin{equation} p^A_j=\sum_{i} p_i \tilde{q}_{ij}+p_0/d_A,
	\end{equation}
	with  $r_j p^A_j$ the probability of result $j$, while 
	\begin{equation}
	\tilde{q}_{ij}=\sum_k q_{ik}|\langle j_A|k_A\rangle|^2\,,
	\end{equation}
	with  $r_j \tilde{q}_{ij}$ the probability of result $j$ in the state  $|\Psi_i\rangle$. The ensuing conditional state at $B$ is 
	\begin{equation} \rho_{B/j}=\frac{1}{p^A_j}\left[\sum_i p_i \tilde{q}_{ij}|j_{i B}\rangle\langle j_{i B}|+p_0\frac{{\mathbbm 1}_{d_B}}{d_A d_B}\right],
	\label{rbj}
	\end{equation}
	where $|j_{i B}\rangle=\frac{1}{\sqrt{\tilde{q}_{ij}}}\sum_k 
	\sqrt{q_{ik}}\langle j_A|k_A\rangle|k_{i B}\rangle$. 
	In terms of  the conditional states (\ref{rbkk}) obtained for  a measurement in the local Schmidt basis,   
	we may rewrite (\ref{rbj}) as 
	\begin{equation}
	\rho_{B/j}=\sum_{k}{\textstyle\frac{|\langle j_A|k_A\rangle|^2 p_k^A}{p_j^A}} U_{jk}\rho_{B/k}U^{\dagger}_{jk},
	\end{equation}
	where $U_{jk}$ is a unitary operator satisfying $U_{jk}|k_{i_k B}\rangle=|j_{i_k B}\rangle$ and $\sum_k\frac{|\langle j_A|k_A\rangle|^2 p_k^A}{p_j^A}=1$. 
	Therefore, concavity and completeness of the measurement operators imply 
	\begin{eqnarray}
	&&S_f(B|A_M)=\sum_j r_j p^A_jS_f(\rho_{B/j})\nonumber\\
	&&\geq\sum_{j,k} r_j |\langle j_A|k_A\rangle|^2 p_k^A  S_f(\rho_{B/k})=\sum_k p_k^AS_f(\rho_{B/k})=S_f(B|A)\,.\nonumber
	\end{eqnarray}
	\qed


\medskip

	\begin{thebibliography}{999}
		
		\bibitem{NC.00} M.A.\ Nielsen, I.L.\ Chuang,
		{\em Quantum Computation and Quantum Information} (Camb.\ Univ. Press,  UK, 2000).
		\bibitem{Modi.12} K. Modi, A. Brodutch, H. Cable, T. Paterek,  V. Vedral, Rev.\
		Mod.\ Phys.\ \textbf{84}, 1655 (2012).
		\bibitem{ABC.16} G.\ Adesso, T. R.\ Bromley,  M.\ Cianciaruso, J. Phys.\ A: Math. Theor. \textbf{49},  473001 (2016).
		\bibitem{FAC.10} A.\ Ferraro, L.\ Aolita, D.\ Cavalcanti, F.M.\ Cucchietti, A.\ Ac\'{\i}n,  Phys.\ Rev.\ A \textbf{81}, 052318 (2010).
		\bibitem{dC.18} G. de Chiara,  A. Sanpera, Rep.\ Prog.\ Phys.\ {\bf 81} 074002 (2018).
		\bibitem{ve.02} F.\ Verstraete, {\em  A study of entanglement in quantum information theory}, PhD Thesis, Katholieke Universiteit Leuven (2002).
		\bibitem{Shi.11} M.\ Shi, F.\ Jiang, C.\ Sun,   J.\ Du, N. J.  Phys. \textbf{13}, 073016 (2011);
		M.\ Shi, W.\ Yang, F.\ Jiang, J.\ Du, J.\ Phys. A: Math. Theor. \textbf{44}, 415304 (2011).
		\bibitem{Shi3.12} M.\ Shi, C.\ Sun, F.\ Jiang, X.\ Yan, J.\ Du,
		Phys.\ Rev.\ A \textbf{85}, 064104 (2012).
		\bibitem{SS.13} K.K.\  Sabapathy,    R.\ Simon,   arXiv:1311.0210 (2013).  
		\bibitem{Jev.14} S.\ Jevtic, M.\ Pusey, D.\ Jennings, T.\ Rudolph, Phys. \ Rev.\ Lett.\  \textbf{113}, 020402 (2014).
		\bibitem{Jev2.14} A.\ Milne, D.\ Jennings, S.\ Jevtic, T.\ Rudolph, Phys.\ Rev.\ A \textbf{90}, 024302 (2014).
		\bibitem{M.14} A.\ Milne, S.\ Jevtic, D.\ Jennings, H. Wiseman, T.\ Rudolph, 
		New J.\ Phys. \textbf{16}, 083017 (2014). Corrigendum: New J. Phys. \textbf{16}, 
		083017 (2014).
		\bibitem{Mc.17} R.\ McCloskey, A.\ Ferraro,  M.\ Paternostro, Phys.\ Rev.\ A \textbf{95}, 012320 (2017).
		\bibitem{GR.14} N.\ Gigena,  R.\ Rossignoli, Phys.\ Rev.\ A \textbf{90}, 042318 (2014).
		\bibitem{Hu.15} X.\ Hu,  H.\ Fan, Phys.\ Rev.\  A \textbf{91}, 022301 (2015).
		\bibitem{Cheng.16}
		S.\ Cheng, A.\ Milne, M.\ J. W.\ Hall,  H.M.\ Wiseman, 
		Phys. Rev. A \textbf{94}, 042105  (2016).
		\bibitem{exp.18} C. Zhang, S.\ Cheng, L.\ Li, Q.\ Liang. B.\ Liu, Y.\ Huang, C.\ Li, G.\ Guo, M.J.W.\ Hall, H.M.\ Wiseman, G.J.\ Pryde, Phys. Rev. Lett. \textbf{122}, 070402 (2019). 
		\bibitem{GRb.14} N.\ Gigena,  R.\ Rossignoli, J. Phys. A: Math. Theor. \textbf{47}, 015302 (2014).
		\bibitem{OZ.01} H.\ Ollivier, W.H.\ Zurek, Phys.\ Rev.\ Lett.\ \textbf{88}, 017901 (2001).
		\bibitem{HV.01} L.\ Henderson,   V.\ Vedral, J.\ Phys.\ A: Math.\ Gen.\ \textbf{34}, 6899 (2001).
		\bibitem{Lec.17} N.\ Canosa, M.\ Cerezo, N.\ Gigena,  R.\ Rossignoli in  {\it Lectures on General Quantum Correlations and their Applications},  F.\ Fanchini, D.\ Soares-Pinto, G.\ Adesso (Eds.),  455  (Springer, 2017).
		\bibitem{Bera.18} A.\ Bera,  T.\ Das, 
		D.\ Sadhukhan, S.S.\ Roy, A.\ Sen(De),  U.\	Sen, Rep.\ Prog.\ Phys.\ {\bf 81}, 024001 (2018).
		\bibitem{KW.02}M.\ Koashi, A.\ Winter, Phys.\ Rev.\ A {\bf 69}, 022309 (2004). 
		\bibitem{Sch1} E. Schr\"odinger, Proc. Camb. Phil. Soc. \textbf{31}, 555 (1935); 
		\textbf{32}, 446 (1936).
		\bibitem{WJD.07} H. M.\ Wiseman, S. J.\ Jones,  A. C.\ Doherty, 
		Phys.\ Rev.\ Lett.\ \textbf{98}, 140402 (2007).
		\bibitem{SNC.14}  P.\ Skrzypczyk, M.\ Navascu\'es,  D.\ Cavalcanti, 
		Phys. Rev. Lett. \textbf{112}, 180404  (2014).
		\bibitem{K.15} I.\ Kogias, A. R.\ Lee, S.\ Ragy,  G.\ Adesso,
		Phys. Rev. Lett.  \textbf{114}, 060403 (2015).
		\bibitem{Ga.15} R.\ Gallego,  L.\ Aolita, Phys. Rev.\ X \textbf{5}, 041008 (2015).
		\bibitem{CS.17} D.\ Cavalcanti, P.\ Skrzypczyk, Rep.\ Prog.\ Phys. \textbf{80}, 024001 (2017).
		\bibitem{Bru.18} T.\ Kriv\'achy, F.\ Fr\"owis, N.\ Brunner, 
		Phys.\ Rev.\ A \textbf{98}, 062111 (2018).
		\bibitem{BL.14} G.B.\ Lemos, J.O. de Almeida, S.P.\ Walborn, 
		P.H.\ Souto Ribeiro,  M.\ Hor-Meyll, Phys.\ Rev.\ A {\bf 89}, 042119 (2014). \bibitem{KW.17}K.H.\ Kagalwala, G.\ Di Giuseppe, A.F.\ Abouraddy, 
		B.E.A.\ Saleh, Nat.\ Comm.\ {\bf 8}, 739 (2017); 
		A.F.\ Abouraddy, G.\ Di Giuseppe, T.M. Yarnall, M.C.\ Teich,  B.E.A.\ Saleh
		Phys.\ Rev.\ A {\bf 86}, 050303(R) (2012).
		\bibitem{LR.19} D.\ Pab\'on, L.\ Reb\'on, S.\ Bordakevich, N.\ Gigena, A.\ Boette, C.\ Iemmi, R.\ Rossignoli, S.\ Ledesma, Phys.\ Rev.\ A {\bf 99}, 062333 (2019). 
		\bibitem{N.05} L.\ Neves, G.\ Lima, J.G.\ Aguirre G\'omez, C.H.\ Monken, C.\ Saavedra,  S.\ P\'adua,
		Phys.\ Rev.\ Lett. \textbf{94}, 100501 (2005).
		\bibitem{S.05} M.N.\  O'Sullivan-Hale, I. A.\ Khan, R.W.\ Boyd,  J.C.\ Howell,  Phys.\ Rev.\ Lett. \textbf{94}, 220501 (2005).
		\bibitem{Pan.12} J.W.\ Pan, Z.B.\ Chen, C.Y.\ Lu, H.\ Weinfurter, A.\ Zeilinger, M.\ Zukowski, Rev.\ Mod.\ Phys.\  \textbf{84}, 777 (2012).
		\bibitem{Sa.11} G.\ Lima, L.\ Neves, R.\ Guzm\'an, E.S.\ G\'omez,W.A.T.\ Nogueira,  A.\ Delgado, A.\ Vargas,  C.\ Saavedra, 
		Opt.\ Exp.\ {\bf 19}, 3542 (2011). 
		\bibitem{Cho.08} J.\ Cho, D.G.\ Angelakis, S.\ Bose  Phys.\ Rev.\ A {\bf 78}, 062338 (2008).
		\bibitem{senko.15} C.\ Senko, P.\  Richerme, J.\ Smith, A.\ Lee, I.\ Cohen, A.\ Retzker, C.\ Monroe,   Phys.\ Rev.\ X \textbf{5}, 021026 (2015).
		\bibitem{spin.18} E.\ Moreno-Pineda, C.\ Godfrin, F.\ Balestro, W.\ Wernsdorfer, M.\ Ruben, Chem.\ Soc.\ Rev., \textbf{47}, 501 (2018).
		\bibitem{F.83} U.\ Fano, Rev.\ Mod.\ Phys.\ \textbf{55}, 855 (1983).
		\bibitem{WW.97} S.\ Hill, W.K.\ Wootters, Phys.\ Rev.\ Lett.\ {\bf 78}, 5022 (1997); 
		W.K.\ Wootters, Phys.\ Rev.\ Lett.\ {\bf 80}, 2245 (1998).
		\bibitem{Pe.96} A.\ Peres, Phys.\ Rev.\ Lett.\ {\bf 77}, 1413 (1996).
		\bibitem{HHH.96} M.\ Horodecki, P.\ Horodecki, R.\ Horodecki, Phys. Lett. A \textbf{ 223}, 1 (1996).
		\bibitem{VW.02} G.\ Vidal,  R.F.\ Werner, Phys.\ Rev.\ A {\bf 65}, 032314 (2002); K.\ Zyczkowski, 
		P.\ Horodecki, A.\ Sanpera,  M.\ Lewenstein, Phys.\ Rev.\ A {\bf 58}, 883 (1998). 
			\bibitem{Huang.14} Y.\ Huang, Phys.\ Rev.\ A \textbf{88}, 014302 (2013).
		\bibitem{W.78} A.\ Wehrl,   Rev.\ Mod.\ Phys.\ \textbf {50}, 221 (1978).
		\bibitem{CR.02}  N.\ Canosa, R.\ Rossignoli, Phys.\ Rev.\ Lett. \textbf{88}, 170401 (2002);  Phys.\ Lett.\ A \textbf{264}, 148 (1999).
		\bibitem{Tsa.88}  C.\ Tsallis, J. Stat. Phys. \textbf{52}, 479 (1988).
		\bibitem{BD.96} C.H.\ Bennett, D.P.\ DiVincenzo, J.A.\ Smolin, W.K.\ Wootters, Phys.\ Rev.\ A {\bf 54}, 3824 (1996).
		\bibitem{NK.12} H.\ Nakazato, T.\ Tanaka, K.\ Yuasa, G.\ Florio, S.\ Pascazio, Phys.\ Rev.\ A {\bf 85}, 042316 (2012); 
		T.\ Tanaka, G.\ Kimura, H.\ Nakazato, Phys.\ Rev.\ A {\bf 87}, 012303 (2013). 
		\bibitem{CRCb.10}  R.\ Rossignoli, N.\ Canosa, L.\ Ciliberti, Phys.\ Rev.\ A \textbf{82}, 052342 (2010). 
		\bibitem{RCM.08} R.\  Rossignoli, N.\ Canosa, J.M.\ Matera,
		Phys. Rev. A \textbf{77},  052322 (2008); Phys. Rev. A \textbf{80},  062325 (2009).
		\bibitem{CRM.10} N.\ Canosa, R.\ Rossignoli, J.M.\ Matera, Phys.\ Rev.\ B \textbf{81},  054415 (2010); 
		A.\ Boette,  R.\ Rossignoli,  N.\ Canosa, J.M.\ Matera, Phys.\ Rev.\ B \textbf{94},  214403 (2016). 
		\bibitem{PRA.12} R.\ Rossignoli, J.M.\ Matera, N.\ Canosa, Phys. Rev. A \textbf{86}, 022104 (2012).
	\end{thebibliography}
\end{document}